\documentclass{emulateapj}

\usepackage{amssymb}

\def\ts     {\thinspace}
\def\kms    {\ts km\ts s$^{-1}$}
\def\etal   {{\rm et\ts al.}}
\def\msol   {$M_{\odot}$}
\def\lsol   {$L_{\odot}$}

\def\aco    {{\rm CO}($J$=1$\to$0)}
\def\bco    {{\rm CO}($J$=2$\to$1)}

\def\eco    {{\rm CO}($J$=5$\to$4)}

\def\pss    {PSS\,J2322+1944}

\slugcomment{draft version \today, Accepted for Publication in the Astrophysical Journal}

\shorttitle{High--Resolution CO(2--1) Imaging of PSS\,J2322+1944}
\shortauthors{Riechers et al.}

\begin{document}

\title{A Molecular Einstein Ring at $z$=4.12:\\ Imaging the Dynamics
  of a Quasar Host Galaxy Through a Cosmic Lens}

\author{Dominik A. Riechers\altaffilmark{1,2,7}, 
  Fabian Walter\altaffilmark{1},  Brendon J. Brewer\altaffilmark{3}, 
  Christopher L. Carilli\altaffilmark{4}, \\
  Geraint F.  Lewis\altaffilmark{3}, Frank Bertoldi\altaffilmark{5}, 
  and Pierre Cox\altaffilmark{6}}

\altaffiltext{1}{Max-Planck-Institut f\"ur Astronomie, K\"onigstuhl
  17, Heidelberg, D-69117, Germany}

\altaffiltext{2}{Astronomy Department, California Institute of
  Technology, MC 105-24, 1200 East California Boulevard, Pasadena, CA
  91125}

\altaffiltext{3}{Institute of Astronomy, School of Physics, A28,
University of Sydney, NSW 2006, Australia}

\altaffiltext{4}{National Radio Astronomy Observatory, PO Box O,
Socorro, NM 87801}

\altaffiltext{5}{Argelander-Institut f\"ur Astronomie, Universit\"at
  Bonn, Auf dem H\"ugel 71, Bonn, D-53121, Germany}

\altaffiltext{6}{Institut de RadioAstronomie Millim\'etrique, 300 Rue
  de la Piscine, Domaine Universitaire, F-38406 Saint Martin
  d'H\`eres, France}

\altaffiltext{7}{Hubble Fellow}

\email{dr@caltech.edu}

\begin{abstract}
  We present high-resolution (0.3$''$) Very Large Array (VLA) imaging
  of the molecular gas in the host galaxy of the high redshift quasar
  PSS\,J2322+1944 ($z=4.12$).  These observations confirm that the
  molecular gas (CO) in the host galaxy of this quasar is lensed into
  a full Einstein ring, and reveal the internal dynamics of the
  molecular gas in this system. The ring has a diameter of
  $\sim$1.5$''$, and thus is sampled over $\sim$20 resolution elements
  by our observations.  Through a model-based lens inversion, we
  recover the velocity gradient of the molecular reservoir in the
  quasar host galaxy of PSS\,J2322+1944. The Einstein ring lens
  configuration enables us to zoom in on the emission and to resolve
  scales down to $\lesssim$1\,kpc. From the model-reconstructed
  source, we find that the molecular gas is distributed on a scale of
  5\,kpc, and has a total mass of $M({\rm H_2})=1.7 \times
  10^{10}\,$\,\msol. A basic estimate of the dynamical mass gives
  $M_{\rm dyn} = 4.4 \times 10^{10}\,\sin^{-2}\,i$\,\msol, that is,
  only $\sim$2.5 times the molecular gas mass, and $\sim$30 times the
  black hole mass (assuming that the dynamical structure is highly
  inclined).  The lens configuration also allows us to tie the optical
  emission to the molecular gas emission, which suggests that the
  active galactic nucleus (AGN) does reside within, but not close to
  the center of the molecular reservoir.  Together with the (at least
  partially) disturbed structure of the CO, this suggests that the
  system is interacting.  Such an interaction, possibly caused by a
  major `wet' merger, may be responsible for both feeding the quasar
  and fueling the massive starburst of 680\,\msol\,yr$^{-1}$ in this
  system, in agreement with recently suggested scenarios of quasar
  activity and galaxy assembly in the early universe.
\end{abstract}


\section{Introduction}

A fundamental aspect in studies of galaxy formation and evolution is
to understand the connection between AGN and starburst activity. The
existence of a physical connection between both processes is suggested
by the finding that present day galaxies show a strong relationship
between the mass of their central supermassive black holes (SMBHs) and
the mass and concentration of their stellar spheroids (Magorrian
\etal\ \citeyear{mag98}; Ferrarese \& Merritt \citeyear{fer00};
Gebhardt \etal\ \citeyear{geb00}; Graham \etal\ \citeyear{gra01}). If
these relations were due to a coevolution of both components during
the early assembly of a galaxy, high-redshift quasars and their
associated host galaxies would be ideal objects to study the active
formation of both SMBHs and bulge stars.

Studies of molecular gas (most commonly rotational transitions of CO),
the prerequisite material that fuels star formation, have become an
important tool to probe distant quasar host galaxies, and revealed
large molecular gas reservoirs of $>$10$^{10}$\,\msol\ in a number of
these sources (see Solomon \& Vanden Bout \citeyear{sv05} for a
general review).  These galaxies typically show huge far-infrared
(FIR) luminosities in excess of 10$^{13}$\,\lsol, which are thought to
be powered by starbursts (and possibly a central AGN; e.g., Omont et
al.\ \citeyear{omo01}; Wang et al.\ \citeyear{wan07}). Observations
of molecular gas trace the regions that can host massive starbursts.
In addition, the velocity structure of molecular line emission has the
potential to constrain the dynamical state of galaxies out to the
earliest epochs.

Rotational molecular line emission typically emerges at FIR to radio
wavelengths, i.e., in the limited wavelength regime where the AGN in
distant quasars does not outshine all other emission.  However, the
cosmological distances of high redshift quasars make it difficult to
resolve the faint emission from their host galaxies a such long
wavelengths.  The physical resolution of such observations is in some
cases boosted by gravitational lenses acting as natural
telescopes. The gravitational lensing effect also magnifies the
observed flux of the background galaxy, in particular for systems in
Einstein ring configurations. Due to the compactness of the AGN,
optical quasars in Einstein ring lens configurations are rare.  Due to
their greater extent, the host galaxies of quasars are much more
likely to cross the inner Einstein ring caustic of a gravitational
lens.

In this paper, we report on high (0.3$''$) angular resolution Very
Large Array (VLA)\footnote{The Very Large Array is a facility of the
National Radio Astronomy Observatory, operated by Associated
Universities, Inc., under a cooperative agreement with the National
Science Foundation.}  observations of CO in the host galaxy of the
$z$=4.12 quasar PSS\,J2322+1944, one of only two known $z$$>$4
galaxies that are both gravitationally lensed and detected in
molecular gas emission (the other being BRI\,0952--0115 at $z$=4.43;
Guilloteau \etal\ \citeyear{gui99}). This galaxy was identified in a
spectroscopic follow-up study of the Palomar Sky Survey (DPOSS;
Djorgovski et al.\ \citeyear{djo00}), and found to be a strongly
lensed optical quasar (S.~G.\ Djorgovski, private communication). It
was subsequently detected in hard X-ray (Vignali et al.\
\citeyear{vig05}), FIR dust (Omont et al.\ \citeyear{omo01}; Isaak et 
al.\ \citeyear{isa02}) and radio continuum emission (Carilli et al.\
\citeyear{car01}), as well as molecular line emission (Cox et al.\
\citeyear{cox02}; Carilli et al.\ \citeyear{car02a}). It follows the 
radio-FIR correlation of star-forming galaxies (Carilli et al.\
\citeyear{car01}; Beelen et al.\ \citeyear{bee06}), indicating that its 
FIR continuum emission is dominated by intense star formation.  In
spite of the fact that this source shows only two unresolved quasar
images in the optical, previous CO observations have shown that the
molecular gas reservoir in its host galaxy is lensed into an Einstein
ring (Carilli \etal\ \citeyear{car03}; hereafter:\ C03). These
observations were also used to derive a first lensing model for this
source.  Based on the dynamical structure revealed by our new, higher
resolution observations of PSS\,J2322+1944, we have developed a new
lensing model, which enables us to reconstruct the velocity gradient
in the spatially resolved gas reservoir.  We use a concordance, flat
$\Lambda$CDM cosmology throughout, with $H_0$=71\,\kms\,Mpc$^{-1}$,
$\Omega_{\rm M}$=0.27, and $\Omega_{\Lambda}$=0.73 (Spergel \etal\
\citeyear{spe03}, \citeyear{spe07}).

\section{Observations}

We observed the \bco\ transition ($\nu_{\rm rest} = 230.53799\,$GHz)
towards \pss\ using the VLA in B configuration between 2006 June 19
and July 3, and in C configuration between 2002 October 21 and
November 15 (these short spacings data were published in the original
study by C03).  The total on-sky integration time in the 11 observing
runs amounts to 70.5\,hr.  At $z=4.119$, the line is redshifted to
45.0351\,GHz (6.66\,mm).  Observations were performed in
fast-switching mode (see, e.g., Carilli \& Holdaway \citeyear{ch99})
using the nearby source J23307+11003 for secondary amplitude and phase
calibration.  Observations were carried out under very good weather
conditions with 26 or 27 antennas. The phase stability in all runs was
excellent (typically $<$20$^\circ$ phase rms for the longest
baselines). The phase coherence was checked by imaging the a
calibrator (J23207+05138) with the same calibration cycle as that used
for the target source.  For primary flux calibration, 3C48 was
observed during each run.  Due to the restrictions of the VLA
correlator, one 50\,MHz intermediate frequency (IF) with seven
6.25\,MHz channels was centered at the \bco\ line frequency, leading
to an effective bandwidth of 43.75\,MHz (corresponding to 291\kms\ at
45.0\,GHz). This encompasses almost the full CO line width as measured
in the \eco\ transition ($\Delta v_{\rm FWHM} = 273 \pm 50$\,\kms, Cox
\etal\ \citeyear{cox02}). Earlier observations set a 2\,$\sigma$ limit
of 150$\mu$Jy on the continuum emission at the line frequency (Carilli
\etal\ \citeyear{car02a}).

For data reduction and analysis, the AIPS package was used. All data
were mapped using the CLEAN algorithm and natural weighting.  The
synthesized clean beam using all data has a size of
0.33$''$$\times$0.30$''$.\footnote{Imaging B array data only gives a
resolution of 0.17$''$$\times$0.15$''$ (natural weighting), or
0.12$''$$\times$0.11$''$ (uniform weighting); however, most of the
emission is outresolved in such maps.}  The final rms over a bandwidth
of 37.5\,MHz (250\,\kms, excluding the noisy edge channel) is
48\,$\mu$Jy beam$^{-1}$.  In addition, seven velocity channel maps
(6.25\,MHz, or 42\,\kms\ each) of the \bco\ emission were created. The
rms after Hanning smoothing is 77\,$\mu$Jy\,beam$^{-1}$.  Convolving
the data to a linear spatial resolution of 0.5$''$ leads to slightly
higher rms values of 49 and 80\,$\mu$Jy beam$^{-1}$ for maps at 250
and 42\,\kms\ velocity resolution.

\section{Results}

\begin{figure}
\epsscale{1.2}
\vspace*{-7mm}

\plotone{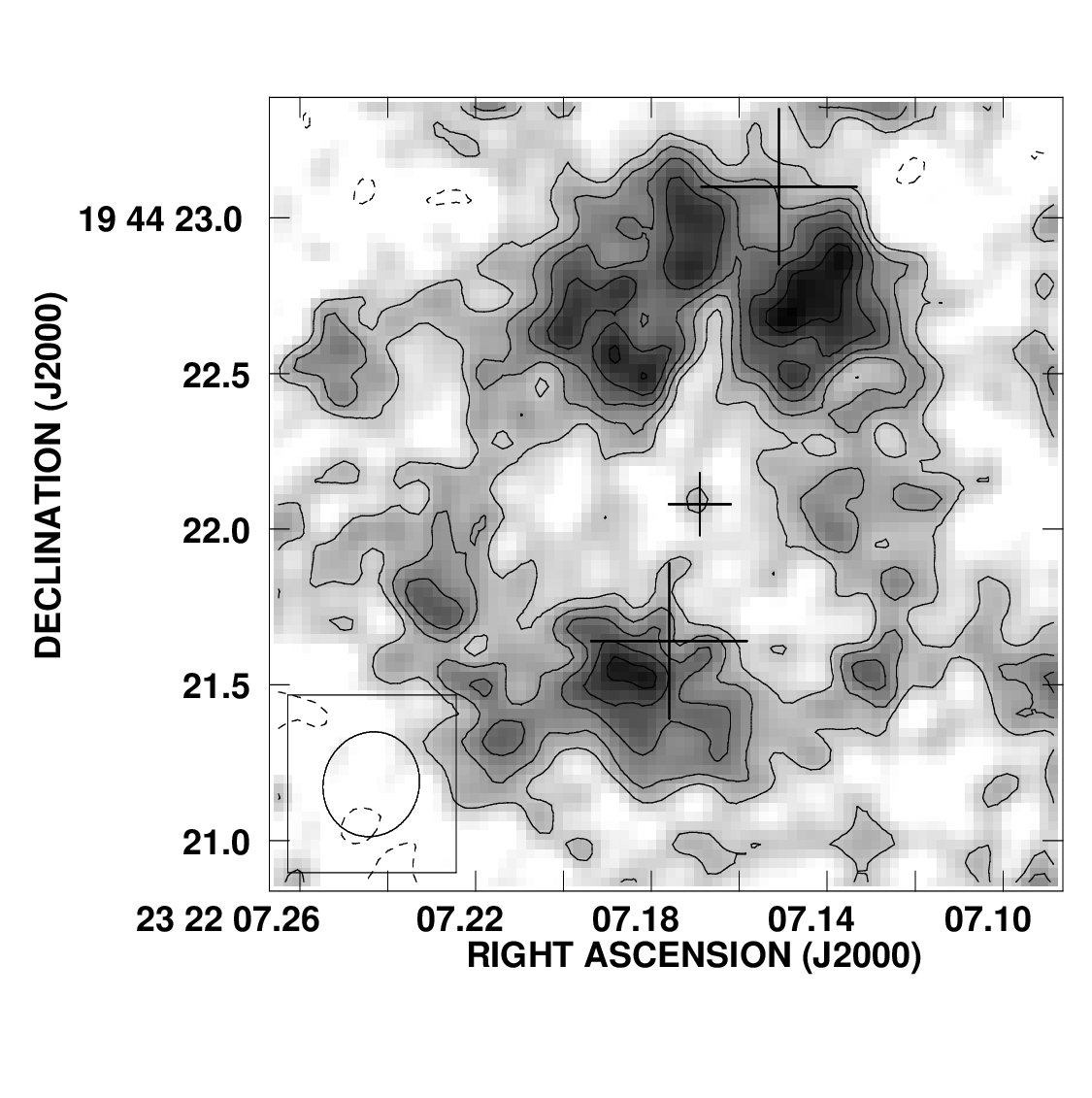}
\vspace*{-12mm}

\caption{VLA map of the \bco\ emission toward \pss\ (integrated over
the central 37.5\,MHz, or 250\,km\,s$^{-1}$).  Contours are shown at
(-3, -2, 1, 2, 3, 4, 5)$\times\sigma$ (1$\sigma = 48\,\mu$Jy
beam$^{-1}$).  The beam size (0.33$''$$\times$0.30$''$) is shown in
the bottom left corner. The large crosses show the positions of the
quasar images at $\lambda_{\rm obs} = 1.6\,\mu$m ($\lambda_{\rm r} =
314$\,nm), and the small cross shows the position of the lensing
galaxy at the same $\lambda$.
\label{f1}}
\end{figure}

In Figure \ref{f1}, the velocity-integrated \bco\ emission over the
central 250\,\kms\ is shown. The emission is clearly resolved over
multiple beams, and extended on a scale of $\sim$1.5$''$.  The
distribution of the gas is reminiscent of a full, almost circular
molecular Einstein ring, consistent with previous indications of such
a structure in lower resolution observations (C03). The emission
varies in intensity along the ring, showing clear substructure.

The apparent surface brightness variations along the ring set strong
constraints on the geometry of the lens configuration.  The brightest
CO peak on the ring can be used to set a lower limit on the intrinsic
brightness temperature of the molecular gas. The peak strength of
280$\pm$48\,$\mu$Jy\,beam$^{-1}$ corresponds to a beam-averaged,
rest-frame brightness temperature of $T_{\rm b}$=8.7$\pm$1.5\,K. This
is by a factor of a few lower than the kinetic gas temperature $T_{\rm
kin}$ as predicted from CO line excitation models of this source
(Riechers et al.\ \citeyear{rie06}; A.~Wei\ss\ et al., in prep.), and
indicates that the substructure of the molecular reservoir is not
fully resolved by the observations.

We derive a spatially integrated \bco\ line peak flux density of 2.50
$\pm$ 0.32\,mJy, which is consistent with that found by Carilli \etal\
(\citeyear{car02a}).  Assuming constant $T_{\rm b}$ between \bco\ and
\aco, and a CO luminosity to H$_2$ mass conversion factor for
ultraluminous infrared galaxies 
[$\alpha$=0.8\,\msol\,(K\,\kms\,pc$^2$)$^{-1}$; Downes \& Solomon
\citeyear{ds98}], this corresponds to a total molecular gas mass of
$M({\rm H_2})=9.0 \times 10^{10}\,\mu_L^{-1}$\,\msol\ (where $\mu_L$
is the lensing magnification factor, see below). Within the
uncertainties, this is in agreement with the value found by Riechers
\etal\ (\citeyear{rie06}) based on the \aco\ luminosity.

\begin{figure}
\vspace{-10mm}

\epsscale{1.2}
\plotone{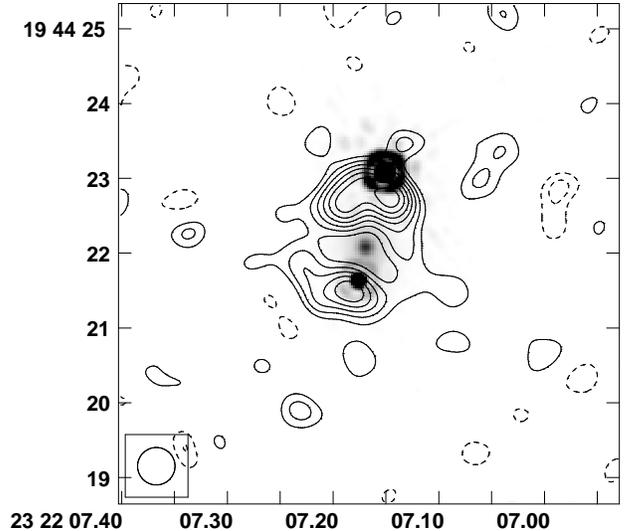}
\vspace*{-17.5mm}

\caption{Contours of the \bco\ emission as shown in Figure \ref{f1},
but convolved to a linear resolution of 0.5$''$ and overlaid on a HST
NICMOS-2 image at $\lambda_{\rm obs} = 1.6\,\mu$m (position crosses in
Figure \ref{f1}). Contours are shown at (-3, -2, 2, 3, 4, 5, 6, 7,
8)$\times\sigma$ (1$\sigma = 49\,\mu$Jy beam$^{-1}$).  Note that the
HST image was not cleaned, and thus shows Airy rings around the
unresolved quasar images (Peng et al.\ \citeyear{pen06}; C.~Y.\ Peng
2008, private communication).
\label{f2}}
\end{figure}

To increase the peak signal--to--noise ratio, Figure \ref{f2} shows
the \bco\ emission convolved to 0.5$''$ linear resolution. This
contour map is shown overlaid on a greyscale image of the (rest-frame)
314\,nm continuum emission of the source, as observed by the Hubble
Space Telescope (HST/NICMOS-2 F160W; image adopted from Peng et al.\
\citeyear{pen06}; C.~Y.\ Peng 2008, private communication). The two
brightest spots in the optical image are unresolved lensed images of
the AGN (henceforth `A' and `B',\footnote{By convention, A is the
brightest image.}  also indicated as large crosses in Figure
\ref{f1}).  The optical emission is likely dominated by the broad-line
region of the AGN, and thus expected to emerge from a compact
circumnuclear pc-scale region.  The quasar images have a brightness
ratio of $b$(B,A)=0.181. They coincide within 0.10$''$ with the
positions measured by Keck at 430\,nm (2.2\,$\mu$m observed frame;
C03), i.e., within the relative astrometric errors. These positions
however are clearly offset from the brightest peaks of the molecular
line emission. The third spot in the optical image is the lensing
galaxy (`G', also indicated as a small cross in Figure \ref{f1}). The
lensing galaxy lies on the axis connecting images A and B, but by more
than a factor of 2 closer to image B. It however lies in the very
center of the molecular Einstein ring, as expected. At the position of
the lensing galaxy, we measure a 45.0\,GHz radio continuum flux of 79
$\pm$ 47\,$\mu$Jy. This corresponds to only 1.7$\sigma$, and thus has
to be considered tentative at best.

In the adopted cosmology, the optical brightness of PSS\,J2322+1944
corresponds to an apparent bolometric luminosity of $L_{\rm bol} = 2.1
\times 10^{14}\,\mu_{L,{\rm opt}}^{-1}$\,\lsol\ (Isaak et al.\
\citeyear{isa02}). Assuming Eddington accretion, this corresponds to
an apparent black hole mass of $M_{\rm BH} = 7.0 \times
10^{9}\,\mu_{L,{\rm opt}}^{-1}$\,\msol. Note that this may be
considered a lower limit, as black holes in $z \sim 4$ quasars are
found to have typical accretion rates of $\dot{M}={L_{\rm bol}/L_{\rm
Edd}} = 0.3-0.4$ (e.g., Shen et al.\ \citeyear{she07}).

\begin{figure*}
\epsscale{1.2}
\vspace{-10mm}

\plotone{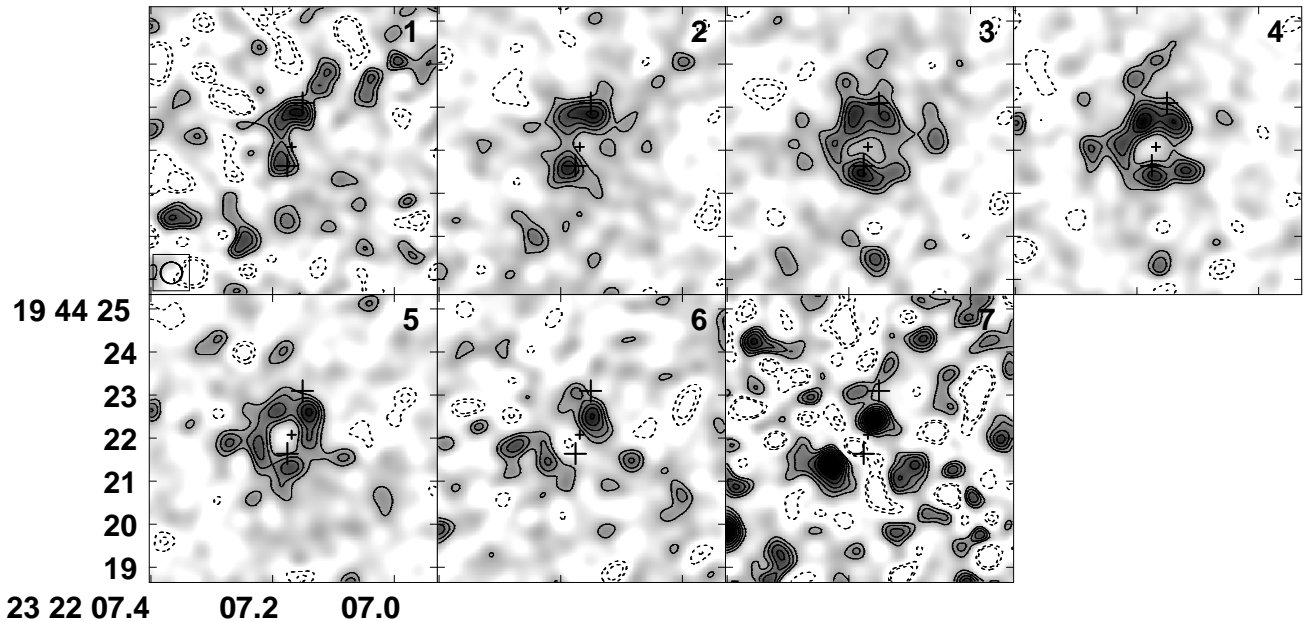}
\vspace{-22mm}

\caption{Channel maps of the \bco\ emission at
0.5$''$ resolution (beam size is shown in the bottom left corner of
the first panel). The same region is shown as in Fig.~\ref{f2}.  One
channel width is 6.25\,MHz, or 42\,\kms\ [frequencies increase with
channel number and are shown at 45016.35, 45022.60, 45028.85,
45035.10, 45041.35, 45047.60, and 45053.85\,MHz (red to blue)].
Contours are shown at (--3, --2, 2, 3, 4, 5, 6)$\times \sigma$
(1$\sigma = 80\,\mu$Jy\,beam$^{-1}$). Note that the noise in channel 7
is intrinsically higher by a factor of $\sqrt{2}$ relative to the
other channels.
\label{f3}}
\end{figure*}

In Figure \ref{f3}, seven 42\,\kms\ wide velocity channels of the
\bco\ emission line are shown (channel 7 was omitted from the data
shown in Figures \ref{f1} and \ref{f2} due to higher noise).  Emission
along the Einstein ring is detected in all channels.  Clearly, the
emission is moving systematically along the ring from the red part of
the \bco\ line to the blue.  Note that there are many peaks of similar
surface brightness (peak fluxes of 450--550\,$\mu$Jy\,beam$^{-1}$, or
$T_{\rm b}$=6--7\,K) that are found at different positions at
different velocities. This indicates, to first order, a dynamical
structure of uniform surface brightness, where the components at
different velocities get projected to different positions in the lens
plane.  This puts us in the unique situation to model the
gravitational lens configuration in this system in detail, and to
recover the intrinsic dynamical structure of the quasar host galaxy in
the background.

\section{Gravitational Lens Inversion}

In a previous attempt to model the gravitational lensing effect toward
the molecular gas reservoir of \pss, C03 adopted a generic model based
on the strongly lensed $z$=0.84 radio galaxy MG\,1131+0456. By
comparing various source configurations in the assumed lens potential
and based on plausibility arguments, a model was found that described
the overall properties of the observed ring structure. To enable us to
describe the intrinsic properties of \pss\ in more detail, we here
present a direct model reconstruction and inversion of the lensing
effect in this system.  This modeling is based on the new CO
observations presented in the previous section.

\subsection{Method: Bayesian Inference}

Due to the remaining observational uncertainties, it is not possible
to derive a {\em unique} solution for the inversion of the
gravitational lens.  We thus explored the parameter space permitted by
the data (and our modeling assumptions) to find the best possible
solution; if some property of the models is consistent throughout this
permitted volume, it can be considered well-constrained at high
confidence.  This parameter study thus follows a Bayesian approach
\citep{gre05}, using 
the Markov Chain Monte Carlo (MCMC) code by \cite{bl06}. This
algorithm does not simply aim at minimizing $\chi^2$ to find the best
model (which may ``overfit'' the data by fitting part of the noise),
but explores the whole range of plausible fits. By doing this, we can
use a large number of parameters (e.g. source pixel values) in our
model without overfitting the noise (since a broad region in parameter
space with higher $\chi^2$ can outweigh a particular solution that has
very low $\chi^2$ if the volume of the parameter space near the very
low $\chi^2$ solution is very small).

The Bayesian analysis thus encodes the phase space of possible source
distributions allowed by the data into a probability distribution
(e.g., Gregory \citeyear{gre05}). The lensed image of the source was
reconstructed based on the integrated CO emission line map and the
structure detected in the velocity channels, and then used to derive a
common model of the lensing galaxy's projected density profile that
reproduces the emission in all channels simultaneously. Due to the
differential structure among the velocity channels, this implies that
the reconstructed source components after lens inversion will be
different for each velocity channel, and will reproduce the velocity
gradient across the source.

In this model description, the unknowns to be inferred from the data
are the seven source profiles $\{s_i\}_{i=1}^7$, one for each velocity
channel, where $s_i$ is shorthand for a large number of (unknown)
pixel values. The unknown lens model parameters are denoted
collectively by $L$. Given observed data $D$, the posterior
distribution for the unknown parameters is proportional to the product
of the prior distribution and the likelihood function:
\begin{equation}
p(\{s_i\}, L | D) \propto p(\{s_i\}, L)p(D | \{s_i\}, L ) .
\end{equation}
Here, $D$ consists of seven extended images (one for each velocity
channel). We make the following standard assumption for $p(D |
\{s_i\}, L )$: Assuming that we know the source and lens properties, we
would predict the observed image to be the lensed, blurred image of
that source, plus additive Gaussian noise:
\begin{equation}\label{likelihood}
p(D | \{s_i\}, L ) \propto \exp\left[-\frac{1}{2}\sum_{i=1}^7\chi_i^2\right],
\end{equation}
where
\begin{equation}\label{chisq}
\chi_i^2 = \sum_{j=1}^{N_{\rm pixels}}\left(\frac{D_{i,j} - M_j(s_i, L)}{\sigma_i}\right)^2.
\end{equation}
Here, $D_{i,j}$ is the $j$'th pixel of the $i$'th image, and $M(s_i,
L)$ is the model image calculated by lensing and blurring the $i$'th
source with the proposed lens model $L$, and $\sigma_i$ is the noise
standard deviation in the $i$'th image estimated from the outer `blank
sky' regions of each image, i.e., distant from any detected structure.

In this study, the lens was parameterized as a singular isothermal
elliptical potential with five free parameters, which are the strength
$b$ of the lens, the ellipticity $q$ of the potential, the central
position $(x_c,y_c)$ of the lensing source, and the angle of
orientation $\theta$ of the projected density profile. For this model,
the lensing potential is
\begin{equation}
\phi(x,y) = b\sqrt{qx'^2 + y'^2/q} \, ,
\end{equation}
where
\begin{equation}
\left(\begin{array}{c}
x' \\
y'
\end{array}\right) =
\left(\begin{array}{cc}
\cos\theta & \sin\theta \\
-\sin\theta & \cos\theta
\end{array}\right)\left(\begin{array}{c}
x-x_c \\
y-y_c
\end{array}\right) .
\end{equation}
The source plane position $(x_s,y_s)$ corresponding to any lens plane position $(x,y)$
is then given by the lens equation:
\begin{eqnarray}
x_s = x - \frac{\partial\phi}{\partial x} \, , \nonumber \\
y_s = y - \frac{\partial\phi}{\partial y} \, .
\end{eqnarray}

The optical position of the lensing galaxy was {\em not} used as an
initial model constraint to allow for a conservative treatment of the
errors. However, the optical data is used in a second step to better
constrain some of the source's intrinsic properties, as described in
more detail below.

\subsection{Application to Interferometric Data}

Interferometer maps are reconstructed from visibility data using a
scale that samples one synthesized interferometer beam (i.e.,
resolution element) with multiple pixels. This means that the noise is
not independent for all pixels, as assumed by
Equation~\ref{chisq}. Correct modelling of correlated noise is
computationally very expensive, and makes the evaluation of the
likelihood (Equation~\ref{likelihood}) very slow. In interferometric
images, the scale of the synthesized beam, or point-spread function
(PSF), usually is the same as the noise correlation scale (depending
on the interferometer baseline weighting function used in the imaging
Fourier transform). A good PSF model thus will allow for a proper, but
faster treatment of the noise properties.

The correlation length scale of the noise was measured at a clear
distance from the detected molecular structure and used as the length
scale for constructing an optimized Gaussian PSF. This length scale
predicts that 1/62 of the pixels are effectively independent at the
scale of the images (0.03$''$\,pixel$^{-1}$). This corresponds to 106
pixels on the scale of the Einstein ring. To not ``overfit'' the image
due to noise, only this fraction of information can be used to
computationally determine the lens properties. This means that MCMC
calculations have to be run at an ``annealing temperature'' of 62, or
alternatively, the $\sigma$'s can be artificially increased by a
factor $\sqrt{62}$ (e.g., Gregory \citeyear{gre05}). The validity of
this short-cut procedure was confirmed by ensuring that the residuals
of the model images have the same statistical properties as the
background in the observed maps.

\subsection{Priors}

The prior distribution for the lens parameters is chosen to be
diffuse, but the final result is independent from this selection. For
the seven unknown sources representing the source distribution in the
velocity channels, we use independent ``massive inference'' priors
\citep{0047}. To generate a random source, 
a moderate number of ``atoms'' of a certain brightness are added to a
blank source plane. These atoms are given a uniform probability
distribution in position and an exponential probability distribution
in flux.  The width of each atom in pixels is chosen at random from
one of three values to allow for the expectation that pixel values
should be correlated with their neighbours. This procedure generates a
random source in which most pixels are dim, and is a more appropriate
prior for astronomical sources than most conventional regularizers
\citep{bl06}.

\subsection{Modeling Results}

\begin{figure*}
\epsscale{1.2}
\plotone{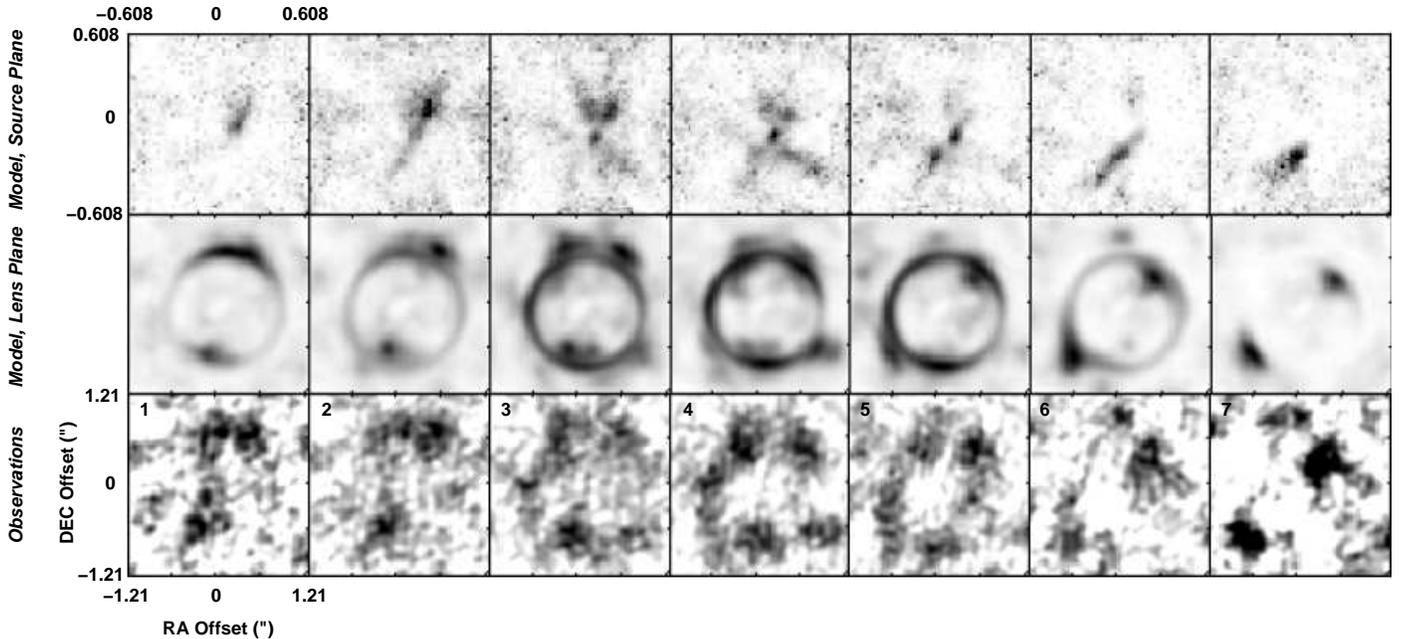}
\caption{Reconstructed model images (posterior mean) of the 42\,\kms\ \bco\ velocity
  channels, and source morphologies after lens inversion. The {\em
  bottom} row shows the observed \bco\ emission in the velocity
  channels as shown in Figure \ref{f3}, but without applying a beam
  convolution (i.e., same resolution as Figure \ref{f1}, thus
  1$\sigma$=77\,$\mu$Jy\,beam$^{-1}$).  The {\em middle} row shows the
  model-reconstructed images of the channel maps in the lens plane
  (same scale as bottom). The {\em top} row shows the reconstructed
  images in the source plane after lens inversion. To account for the
  magnification of physical scales (and thus in effective spatial
  resolution), the model reconstruction in the source plane (same
  region as shown in Figure \ref{f4}; scale in arcsec given on top) is
  shown zoomed-in by a factor of 2 relative to the lens plane.
  \label{f5}}
\end{figure*}

\subsubsection{Lens Parameters}

From the MCMC run, we find that the lens has a strength $b$ =
0.745$''$ $\pm$ 0.014$''$, an ellipticity $q$ = 0.969 $\pm$ 0.014, and
a position ($x_c$ = 0.074$''$ $\pm$ 0.024$''$; $y_c$ = --0.109$''$
$\pm$ 0.029$''$) [coordinates are relative to the center of the model
images at $\alpha$=$23^{\rm h}22^{\rm m}07^{\rm s}.176$,
$\delta$=$+19^\circ44'22''.16$]. Due to the fact that $q \simeq 1$,
the lens potential is close to circular, as is the projected lens
density profile. Even though the observed position of the lens
($x_c^{\rm obs}$ = 0.105$''$; $y_c^{\rm obs}$ = --0.080$''$) was not
taken as an input parameter, the model naturally reproduces its
position within the errors. The posterior probability distribution for
$\theta$ is bimodal; with a probability of 62\% (38\%), the angle of
orientation of the potential is $\theta$ = 109.2$^\circ$ $\pm$
7.1$^\circ$ (62.7$^\circ$ $\pm$ 6.5$^\circ$). The lens parameters are
constrained well by the molecular data alone, so the marginal
posterior distributions are close to Gaussian (except for the bimodal
$\theta$ distribution, which is well approximated by a mixture of two
Gaussians). Thus, all of the estimates and uncertainties quoted above
are of the form (mean $\pm$ standard deviation).

The lensing model described above was derived based on the
distribution of the lensed CO emission only to allow for a
conservative treatment of the errors, and to avoid a main systematic
source of error:\ the remaining astrometric uncertainties between the
optical and radio reference frames. The results of our study indicate
that the optical position of the lens is reproduced well, and thus,
that the astrometric offset appears to be small. Due to the fact that
gravitational lensing is achromatic, we thus can use the model derived
based on the distribution of the lensed CO emission only to also
constrain the intrinsic optical properties of the source. We also can
use the positions of the lensed images of the quasar in the optical to
further constrain the allowed parameter space, and to estimate the
position of the AGN within the deprojected molecular gas reservoir.

Based on the sample of lenses that fit the molecular line data, the
optical positions of the quasar were ray-traced back into the source
plane. As the optical emission of the source is compact, only those
models that map both quasar images onto the same position in the
source plane within the errors can be considered valid. We thus
discarded all models that did not fulfill this extra criterion.

We find that this extra constraint only marginally changes previous
results for the strength, ellipticity, and position of the
lens. However, it does impact the solution for the angle of
orientation of the potential, and supports $\theta \sim 109^\circ$.
As the source reconstructions presented in this section are not
strongly sensitive to the extra constraint, only results (and the more
conservative errors implied) from simulations produced without using
the optical data are shown unless stated otherwise.

\subsubsection{Source Profiles}

For the seven source profiles in the different velocity channels, a
``best'' estimate was obtained by taking the average of all sources
encountered by the MCMC run. This gives the posterior mean for each
source, which is an optimal estimate, as it minimizes the expected
squared error. As no unique solution exists for the lens inversion,
the full source sample has a certain diversity.  However, by taking
the mean, only reproducible features are retained from the full
sample.  Due to the fact that both the resolution and sensitivity of
the observations are finite, the resulting surface brightness profiles
are expected to be smooth on a certain critical scale. Substructure
may appear on smaller scales, but will be smoothed out in the final
model due to the larger uncertainites involved in reproducing such
small structures across the permitted volume in parameter space.

In Figure \ref{f5}, the model-reconstructed (posterior mean) molecular
gas distribution in the seven CO velocity channels ({\em columns}) are
shown in the lens plane ({\em middle row}), and after lens inversion
({\em top row}), together with the observations (not convolved, i.e.,
at full spatial resolution; {\em bottom row}).  Since the lens model
is common to all velocity channels, the changing distribution of
molecular gas between the velocity channels is due to intrinsic
velocity structure in the background source (note the difference in
scale between the lensed and the unlensed source).

\begin{figure*}
\epsscale{1.2}
\vspace{-3mm}

\plotone{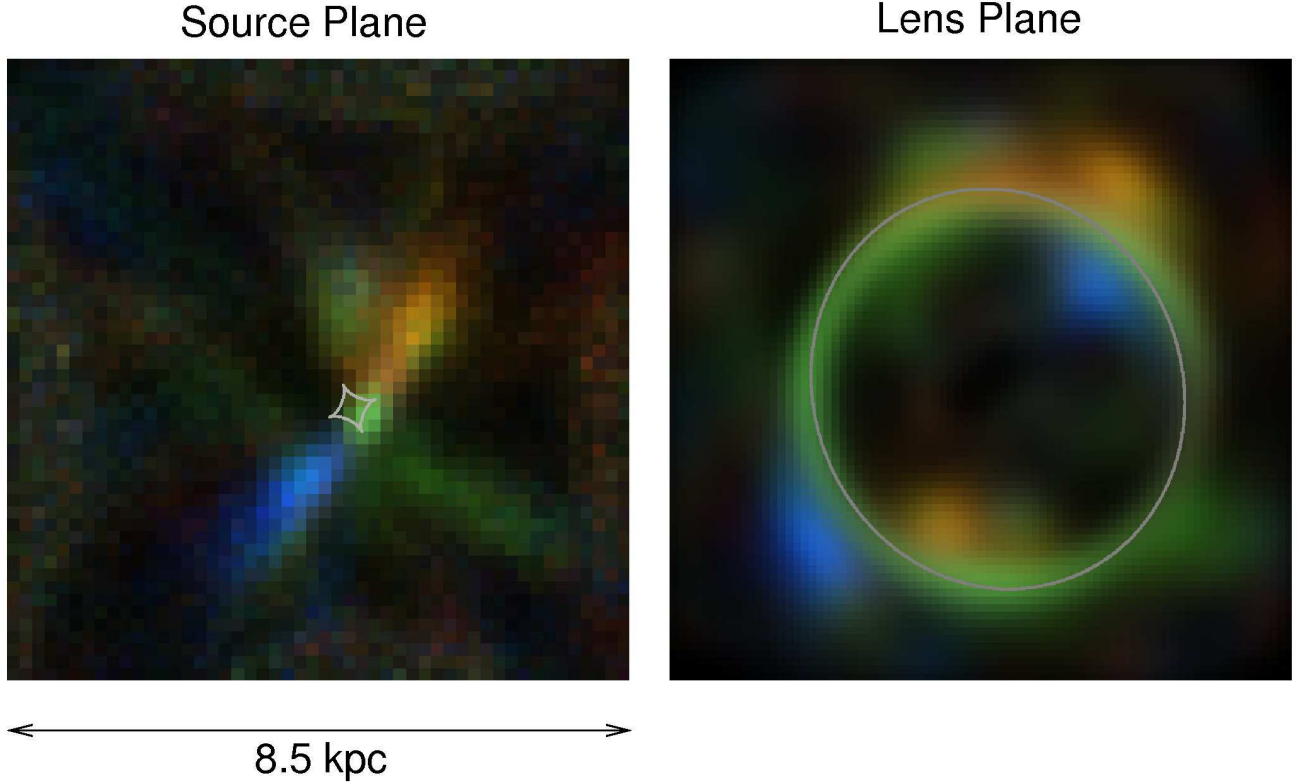}
\vspace{-5.5mm}

\caption{RGB composite color map of the model-reconstructed \bco\
  velocity channels shown in Figure \ref{f5}, with three colors
  encoding the velocity range of the emission [orange: redshifted
  (channels 1-2), green: central (channels 3-5), blue: blueshifted
  (channels 6-7)]. The grey lines indicate the caustics.  The bar at
  the bottom indicates the scale in the source plane.  {\em Right}:
  Lensed, blurred image of the Einstein ring in the lens plane. {\em
  Left}: Image of the model-reconstructed quasar host galaxy in the
  source plane.  \label{f7}}

\end{figure*}

In Figure \ref{f7}, a color composite map of the modeled distribution
of the CO in the velocity channels is shown.  Orange corresponds to
the redshifted part of the emission line, green corresponds to the
central part, and blue corresponds to the blueshifted part. In the
{\em right} panel, the image in the lens plane is displayed, and in
the {\em left} panel, the image in the source plane is shown after
lens inversion. A clear velocity gradient is seen in both images. The
brightest part of the emission shows a systemic velocity structure
that would be in agreement with a highly inclined, almost edge-on
rotating disk of $\sim$2.5\,kpc radius within the uncertainties of the
data and modeling.  This structure encloses a dynamical mass of
$M_{\rm dyn} = 4.4 \times 10^{10}\,\sin^{-2}\,i$\,\msol\ (assuming a
linewidth of 273\,\kms\ as given above), corresponding to only 2.5
times the lensing-corrected molecular gas mass (for highly inclined
dynamical structures).  However, there is also indication for a
second, possibly tidal component (blue/green).  This structure is
slightly bent, and also has an extent of almost 5\,kpc, similar to the
brighter component.  Each component extends over at least 4-5
resolution elements in the source plane, and thus is clearly detected
and resolved.

The remaining uncertainties in the source reconstruction can be
quantified by examining individual MCMC samples to calculate a
probabilistic description of different properties of the source. A
manual comparison of models considered to be ``good'' by our
statistical analysis however shows that all share the same principal
features, and only show significant differences in substructure on
scales that are close to the resolution and detection limit, as
expected. It thus is plausible that the overall structure and velocity
gradient along the reconstructed source are real.

\begin{figure}
\epsscale{1.4}
\plotone{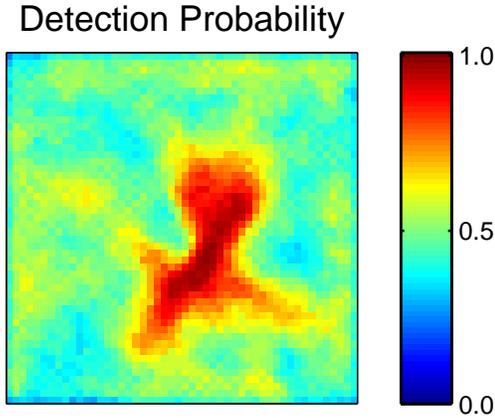}
\caption{Map of the detection probability $P$(pixel $>$ 0)
in the source plane after lens inversion, as derived from the full
MCMC parameter study. A region of 1.21$''$$\times$1.21$''$ size is
shown.
\label{f4}}
\end{figure}

To quantify the uncertainties in the structure of the molecular gas
reservoir of \pss\ in the source plane, the posterior probability $P$
that the flux in each pixel is nonzero is shown in
Figure~\ref{f4}. The detection probability is shown over the full
velocity range (i.e., summed up over all velocity channels). Due to
the fact that observational data contains noise (i.e., finite positive
or negative flux in every pixel of any velocity channel), the
detection probability is nonzero in every pixel.  Thus, a detection
probability below a certain threshold has to be considered
``background''. The detection probability is consistently
significantly higher than the background along the structures
predicted by the ``best model'' (see Figures \ref{f5} and
\ref{f7}).

\begin{figure}
\epsscale{1.2}
\plotone{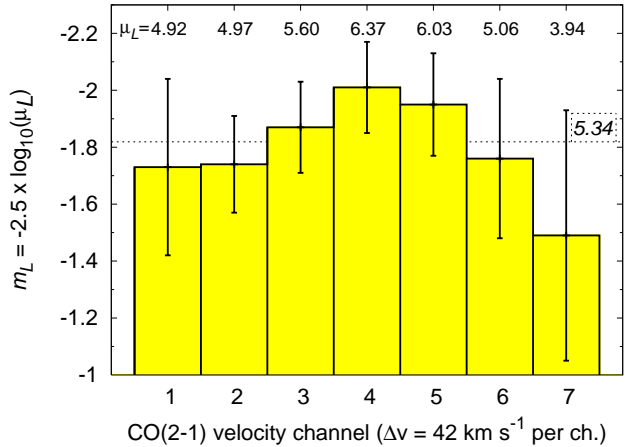}
\caption{Model-predicted differential gravitational magnification
between the CO velocity channels shown in Figures \ref{f3} and
\ref{f5}.  The magnification is shown in magnitudes (i.e., no
magnification corresponds to $m_L=0$). The error bars include the
modeling uncertainties. The numbers on top indicate the magnification
in each velocity channel. The dashed line indicates the total
magnification of $\mu_L$=5.34 in the integrated emission line map.
\label{f6}}
\end{figure}

The overall lensing magnification of the CO emission as predicted by
the posterior distribution for the lens model is $\mu_L = 5.34 \pm
0.34$. Figure \ref{f6} shows the differential flux magnification
between the velocity channels, estimated by calculating flux(lensed
image)/flux(source) for the MCMC samples of each velocity
channel. Although consistent with unity within the modeling
uncertainties, there is tentative evidence that the flux is more
gravitationally enhanced in the line center than in the outer
channels.  If real, this would suggest that the profile of the \bco\
emission line is distorted by the lensing effect.

The model-predicted lensing magnification of the 314\,nm continuum
emission from the AGN as derived from models based on both optical and
CO data is $\mu_L^{\rm opt} = 4.7 \pm 0.4$. Excluding the optical
constraints would result in slightly larger errors: $\mu_L^{\rm opt} =
4.8 \pm 0.6$.  The overlay in Figure \ref{f2} suggests that the
optical AGN is offset from the peak position of the molecular line
emission.  In the red CO line wing, the northern peak of the Einstein
ring is brighter (as is the case for the two optical images), while in
the blue linewing, the southern peak is brighter. This, together with
the optical quasar image positions relative to the lens position,
suggests that the AGN is located close to the orange peak of the
reconstructed molecular gas distribution shown in Figure
\ref{f7}. This finding is supported by the model-based inversion of
the optical and CO data, although the determination of the exact
location within that part of the molecular structure is limited by the
remaining model uncertainties and the limited accuracy of the relative
astrometry of the optical and radio data.  If the AGN is indeed
located in the upper part of the disklike structure rather than in the
center, one may speculate that the spatial and velocity structure of
the reconstructed source is more likely to be due to interaction than
due to a rotating disk.

\section{Discussion}

We have imaged and modeled a molecular Einstein ring of a galaxy at
$z$=4.12. Our high resolution \bco\ maps of the lensed quasar host
galaxy of PSS\,J2322+1944 (a double image optical quasar) reveal
spatially resolved structure that shows a clear velocity gradient in
the CO emission line. By performing a model-based lens inversion of
the Einstein ring that is consistent with the data, we are able to
reconstruct the velocity structure of this distant quasar host galaxy.
The gravitational lensing effect acts as a natural telescope, and
allows us to zoom in on the molecular gas reservoir down to linear
scales of only $\lesssim$1\,kpc, sufficient to reveal velocity
structure over almost 10 resolution elements in the source plane. Our
novel modeling of this system reveals how the molecular gas crosses
the central caustic (causing the appearance of the Einstein ring)
moving from the redshifted to the blueshifted molecular emission. We
also find evidence that the optical quasar may be associated with the
redshifted part of the molecular reservoir.  The full reservoir has a
mass of $M({\rm H_2})=1.7 \times 10^{10}\,$\,\msol\ (corrected for
lensing magnification).  The molecular gas mass alone could account
for almost half of the dynamical mass in this system if the galaxy
were to be seen close to edge-on ($M_{\rm dyn}$\,sin$^2\,i$/$M({\rm
H_2})$ $\simeq$ 2.5).  Due to the large spatial extent of the CO
emission, and due to the fact that the AGN is probably largely offset
from the center of the reservoir, we conclude that the molecular gas
and dust are likely dominantly heated by star formation.

From the FIR luminosity of the source in the adopted cosmology
($L_{\rm FIR} = 2.4 \times 10^{13}\,{\mu_L}^{-1}$\,\lsol; Cox \etal\
\citeyear{cox02}), we derive a star formation rate
\footnote{Assuming
SFR=1.5$\times$10$^{-10}\,L_{\rm FIR}$(\msol\,yr$^{-1}$/\lsol)
(Kennicutt \citeyear{ken98a}). The dust temperature and spectral index
of \pss\ indicate that the FIR continuum emission is dominated by star
formation, in agreement with the finding that the source follows the
radio-FIR correlation for star-forming galaxies (Beelen et al.\
\citeyear{bee06}).} (SFR) of 680\,\msol\,yr$^{-1}$.
At least part of the CO emission of the reconstructed source does not
appear to follow a systemic trend in velocity.  In this picture, this
structure may be due to interaction, possibly caused by a major
merger.  Such an event could both feed the AGN and fuel the starburst,
and thus be responsible for the coeval assembly of a supermassive
black hole and the stellar bulge in this system.  Future observations
of the FIR continuum at comparable spatial resolution may shed more
light on this situation.

Motivated by these results, the dynamical mass derived from the
molecular line observations in the host galaxy of the $z$=4.12 quasar
\pss\ can be used in an attempt to constrain the relationship between
the central SMBH mass and the stellar bulge mass ($M_{\rm
BH}$--$M_{\rm bulge}$) in high-$z$ AGN galaxies. Such a relation has
been proposed for different types of galaxies in the local universe,
and appears to hold over more than three orders of magnitude in SMBH
mass, essentially independent of galaxy type (predicting $M_{\rm
bulge} \simeq 700\,M_{\rm BH}$, e.g., Kormendy \& Gebhardt
\citeyear{kg01}).
Assuming the optical lensing factor of $\mu_L^{\rm opt} = 4.7$ derived
from our model of \pss\ and the Eddington limit derived in Section 3
gives $M_{\rm BH} = 1.5 \times 10^9\,$\msol. With the further
assumption that the dynamical molecular structure is seen close to
edge-on, and that it traces a major fraction of the gravitational
potential that hosts the stellar bulge, we thus find that $M_{\rm
bulge} \simeq 30\,M_{\rm BH}$ (subtracting out the black hole and gas
masses would even give $M_{\rm bulge} < 20\,M_{\rm BH}$). This value
is by more than an order of magnitude offset from the local $M_{\rm
BH}$--$M_{\rm bulge}$ relation, but in good agreement with results
obtained for other high-$z$ quasars, which show similar or larger
offsets (e.g., Walter et al.\ \citeyear{wal04}; Wei\ss\ et al.\
\citeyear{wei07}; Riechers et al.\ \citeyear{rie07}; see also Shields
et al.\ \citeyear{shi06}).

In the case of \pss, one may attempt to account for this offset by
assuming that the dynamical structure is seen close to face-on (i.e.,
$i < 12^\circ$).  This would however predict a large intrinsic CO
linewidth ($\Delta v_{\rm FWHM} > 1300$\,\kms ). Such a linewidth
would significantly exceed the velocity dispersions observed in the
spheroids of massive present day elliptical galaxies (which \pss\ will
likely eveolve into), but is of the same order of magnitude as those
observed toward some high-$z$ submillimeter galaxies (e.g., Carilli \&
Wang \citeyear{cw06}).  Such large molecular linewidths are consistent
with those predicted by simulations of the hierarchical buildup of
massive quasar host galaxies at high redshift (e.g., Narayanan et al.\
\citeyear{nar07}), and thus compatible with the possible merger nature
of \pss. However, in interacting or merging systems, molecular lines
are likely more broad due to the fact that the dynamical molecular
structure is not fully virialized yet. The molecular line widths in
these galaxies thus may be by a factor of a few higher than the actual
virial velocity of the host halo, which would lead to an
overprediction of the bulge mass.  It thus appears difficult to
explain the offset from the local $M_{\rm BH}$--$M_{\rm bulge}$
relation by simply assuming a small inclination angle toward the line
of sight.  Also, assuming that the AGN accretes at sub-Eddington
rates, and/or taking into account that more than a third of the
dynamical mass derived above is likely not stellar, but accounted for
by gas, dust ($<$1\% of the gas mass), and the black hole further
increases this offset. Together with previous such examples, our
results for \pss\ thus appear to indicate that the black holes in
massive galaxies at high redshift assemble earlier than a large
fraction of their stellar bulges.

The observations and modeling presented herein demonstrate the power
of spatially and dynamically resolved molecular gas studies in
strongly lensed, distant AGN-starburst systems to provide direct
evidence for the scenarios of quasar activity and galaxy assembly in
the early universe as suggested by recent cosmological simulations
(e.g., Springel \etal\ \citeyear{spr05}). The boost in line intensity
and spatial resolution provided by Einstein ring lens configurations
are currently the only means by which to probe the dynamical structure
of the most distant star-forming galaxies at (sub-)kiloparsec
resolution. Such observations provide an important foundation for
future observations of molecular gas and dust in the early universe
with the Atacama Large Millimeter/submillimeter Array (ALMA), which
will be able to probe more typical galaxy populations at high redshift
to comparable and higher physical resolution, even without the aid of
gravitational lensing.

\acknowledgments 
The authors would like to thank Dennis Downes for helpful discussions,
and Chien Y.~Peng for providing an HST image of PSS\,J2322+1944.  DR
acknowledges support from the Deutsche Forschungsgemeinschaft (DFG)
through Priority Program 1177, and from NASA through Hubble Fellowship
grant HST-HF-01212.01-A awarded by the Space Telescope Science
Institute, which is operated by the Association of Universities for
Research in Astronomy, Inc., for NASA, under contract NAS 5-26555. CC
acknowledges support from the Max-Planck-Gesellschaft and the
Alexander von Humboldt-Stiftung through the Max-Planck-Forschungspreis
2005. We thank the anonymous referee for a thorough reading of the
manuscript.

\end{document}